\documentstyle[12pt,aaspp4]{article}
                                \begin{document}
                                \title{
Seeing Galaxies Through Thick and Thin:\\
I. Optical Opacity Measures in Overlapping Galaxies
                                }

\vfil								

                                \author{
RAYMOND E. WHITE III$^{1,2,3}$ AND WILLIAM C. KEEL$^{1,2,3}$
                                }
				\affil{
Department of Physics \& Astronomy, University of Alabama, Tuscaloosa, 
AL 35487-0324   
                                }

				\author{
AND
                                }
				\author{
CHRISTOPHER J. CONSELICE$^{2,4,5}$
                                }
                                 \affil{
Department of Astronomy \& Astrophysics, University of Chicago
                                }
                                \vfil				
\noindent
$^1$Visiting astronomer, Kitt Peak National Observatory, National Optical
    Astronomy Observatories,  operated by AURA, Inc., under cooperative
    agreement with the National Science Foundation.

\noindent
$^2$Visiting astronomer, Cerro Tololo Inter-American Observatory, 
    National Optical Astronomy Observatories, likewise operated by AURA, Inc., 
    under cooperative agreement with the National Science Foundation.

\noindent
$^3$Visiting astronomer, Lowell Observatory, Flagstaff, Arizona

\noindent
$^4$NSF REU summer student at University of Alabama

\noindent
$^5$present address: Department of Astronomy, University of Wisconsin

\eject								

                                \begin{abstract}
We describe the use of partially overlapping galaxies to provide
direct measurements of the effective absorption in galaxy disks, 
independent of assumptions about internal disk structure.
The non-overlapping parts of the galaxies and symmetry considerations are 
used to reconstruct, via differential photometry, how much background galaxy 
light is lost in passing through the foreground disks. 
Extensive catalog searches yield $\sim15-25$ nearby galaxy pairs suitable 
for varying degrees of our analysis;  
ten of the best such examples are presented here.
From these pairs, we find that interarm extinction is modest, 
declining from $A_B \sim 1$ magnitude at $0.3 R_{25}^B$ to 
essentially zero by $R_{25}^B$; 
the interarm dust has a scale length consistent with that of the 
disk starlight.
In contrast, dust in spiral arms and resonance rings may be optically thick 
($A_B > 2$) at virtually any radius. 
Some disks have flatter extinction curves than the Galaxy, 
with $A_B/A_I \approx 1.6$; this is probably the signature of clumpy 
dust distributions.
Even though typical spirals are not optically thick throughout their
disks, where they {\it are} optically thick is correlated with
where they are most luminous: in spiral arms and inner disks.
This correlation between absorption and emission regions may account for 
their apparent surface brightness being only mildly dependent on inclination, 
erroneously indicating that spirals are generally optically thick.
Taken as an ensemble, the opacities of spiral galaxies may be just
great enough to significantly affect QSO counts, though not enough
to cause their high redshift cutoff.

                                \end{abstract}

                                \clearpage

                                \section{
Introduction
                                }

Interest in the dust content of spiral disks, particularly in its role as
a source of opacity in ``typical" galaxies,
has been revived by several recent studies.
Different aspects of this problem have been clarified by a 
variety of observational approaches: 

The inclination--surface-brightness test is one of the oldest methods used 
to determine whether spiral galaxies are largely transparent or opaque 
(Holmberg 1958) and this test is still being refined (Valentijn 1990; 
Burstein et al 1991).  An opaque spiral disk would have the same surface 
brightness regardless of its inclination, while a transparent 
disk would have a higher surface brightness when edge-on than face-on.  
Applying this test to a sample of galaxies drawn from the $ESO-LV$ catalog, 
Valentijn (1990) found spirals to be largely opaque. 
This seems counterintuitive for two 
reasons: 1) if spirals are opaque, then the galaxy survey used to
deduce this would have been difficult to obtain, since we live in
a spiral galaxy; 2) there are also well-known examples of distant 
objects (galaxies, quasars, $etc.$) seen through foreground spiral 
galaxies other than the Milky Way.  
In a statistical reassessment of Valentijn's (1990) work,
Burstein $et$ $al.$ (1991) concluded that Valentijn got the right answer
for the wrong reason, maintaining that the result was a product of sample 
selection effects.  Using a sample claimed to be less subject to such selection 
effects, Burstein $et$ $al.$ (1991) nonetheless found that spirals are
optically thick (although not $opaque$, per se).
Most recently, however, Burstein et al. (1995) now finds in an expanded sample
that spirals are $not$ so optically thick after all, since the new surface 
brightness sample exhibits a mild inclination dependence.
Huizinga (1994) has suggested that the Valentijn (1990) result was confounded 
by the presence of bulge systems in the sample, the surface brightnesses of 
bulges being inherently more inclination-independent than those of spiral disks.
However, it is obvious from illustrations in Valentijn (1990) that there 
is a large variance in spiral surface brightnesses at a $given$ inclination, 
which would mask a mild trend of surface brightness with inclination.  
This may explain why this classical method is still of interest --- it gives 
ambiguous results!

Another approach to determining dust opacity in spiral disks is to
compare radiative-transfer models with observed color and surface-brightness 
data (Disney et al 1989; Davies 1990; Witt, Thronson \& Capuano 1992).
While the traditional interpretation is that we see most of the starlight,
free of much reddening or obscuration (Holmberg 1958; 
de Vaucouleurs et al. 1991), these radiative transfer 
studies show that the observed color and surface-brightness
data do {\it not} require low internal extinction, and can be modeled
just as well by very dust-rich systems, in which the optical light
is dominated by the small fraction of least-obscured stars.

Observing the kinematics of edge-on galaxies at various wavelengths offers 
another means for assessing absorption in disks (Bosma et al 1992). 
Using a 21-cm H I rotation curve as a template, one can determine how deeply 
an observed rotation curve at some optical or near-infrared band probes into 
the disk. The two galaxies observed by Bosma et al (1992) were shown to be 
largely transparent.

When seen behind foreground galaxies, the colors of QSOs or the Balmer 
decrements of H II regions can also 
be used to provide crude limits on foreground reddening. 
James \& Puxley (1993) analyzed the Balmer decrements of two H II regions 
projected behind the inner disk of NGC 3314, the foreground galaxy of an 
exactly superposed pair (first analyzed by Keel 1983); they  
found extinctions of $A_B\approx1.7$ magnitudes.
However, when applied to individual QSOs or H II regions, 
this technique selects against high opacity regions, which would
completely obscure small background objects.

Comparing images at widely disparate wavelengths such as $B$ and $K$ can
also be used to determine the intrinsic absorption of spiral disks
(Block et al 1994). A control image is taken in a band at wavelengths long
enough to be minimally affected by dust and compared to an image taken at 
shorter wavelengths. 
Some of the structure in the resulting color map can be attributed
to the reddening effects of dust. Block et al (1994) find that dust
distributions tend to be very patchy and concentrated along spiral arms.
However, this measurement is also sensitive to stellar
population gradients and to the vertical 
distribution of absorbing material, since material at large $z$-distances will 
be most effective at absorbing the overall disk radiation (a fact used
by Elmegreen 1980 to model the scale heights of various disk
constituents as well as the dust in spirals). Furthermore, because
the three-dimensional geometry is not known in detail, effects of
scattering are important in the interpretation (as seen in the recent
claim by Block et al. 1996 that scattering can serve to hide grand-design
spiral patterns in apparently flocculent spirals).

Inspired initially by the work of Valentijn (1990), we initiated
a program to determine the 
opacity of spiral disks $directly$, rather than statistically, by imaging 
foreground spirals partially projected against background galaxies.  
The non-overlapping regions of a partially overlapping galaxy pair can be used 
to reconstruct, using purely differential photometry, how much light from the 
background galaxy is lost in passing through the foreground galaxy in the 
region of overlap.  Initial results are presented in White \& Keel (1992),
Keel \& White (1995), and White, Keel \& Conselice (1996).

                                \section{
Methodology                
                                }
                                \subsection{
Constructing Opacity Maps                
                                }
Our technique for constructing disk opacity maps using purely differential 
photometry is illustrated in Figure 1.  The ideal case consists of a 
foreground disk (spiral) galaxy half-projected against half of a 
similarly-sized background elliptical galaxy.  
For the sake of illustration, the (unobscured) 
surface brightness of each galaxy is taken to be constant, with
$F$ and $B$ being the actual surface brightness values of
the foreground disk and background elliptical in the overlap region, 
and $\tau$ is the optical depth in the disk.
The observed surface brightness in the overlap region is then 
$\langle F+Be^{-\tau}\rangle$,
where brackets are used to emphasize that this whole quantity 
is the observable in the overlap region and cannot be directly
decomposed into its constituent components.
We use symmetric counterparts from the non-overlapping regions of the 
two galaxies to $estimate$ $F$ and $B$ and denote the estimates 
as $F^\prime$ and $B^\prime$.
We can then construct an estimate of
the optical depth, denoted $\tau^\prime$, as follows:
$$e^{-\tau^\prime}= { {\langle F+Be^{-\tau}\rangle - F^\prime} 
\over B^\prime }.
\eqno(1)$$
Here the estimate of the foreground spiral's surface brightness, $F^\prime$,
is first subtracted from the surface brightness of the overlap region,
$\langle F+Be^{-\tau}\rangle$; this result is then divided by the estimate of
the background elliptical's surface brightness, $E^\prime$.
This creates a map of $e^{-\tau^\prime}$ in the overlap region.

Although it is impossible to actually do so strictly from observable quantities,
it is $formally$ useful to ``break" $\langle F+Be^{-\tau}\rangle$, the observed 
surface brightness in the overlap region, into its constituent parts to assess 
the systematic errors of the above construction:

$$e^{-\tau^\prime}\cong { {(F-F^\prime)}\over B^\prime}+{B\over 
B^\prime}e^{-\tau}.
\eqno(2)$$
The systematic errors induced by departures from symmetry can be estimated 
from the non-overlapping parts of the galaxies.
Note that when the background galaxy has substantially higher surface 
brightness than the foreground galaxy ($B^\prime \gg F,F^\prime$), the 
estimate of $\tau$ is particularly insensitive to systematic errors induced by 
asymmetries in the foreground spiral. 
In this case,
$$ e^{-\tau^\prime}\approx {B\over B^\prime}e^{-\tau},
\eqno(3)$$
with $B/B^\prime$ being especially close to unity for most ellipticals and S0s.
Also, a lower limit to $\tau^\prime$ is provided by simply dividing the overlap 
region by the symmetric counterpart of the background galaxy and
neglecting to scrape off the emission from the foreground galaxy:
$$e^{-\tau^\prime} <  { {\langle F+Be^{-\tau}\rangle} \over B^\prime } \qquad
\Rightarrow \qquad \tau^\prime > - {\rm ln} {\langle F+Be^{-\tau}\rangle 
\over B^\prime } .
\eqno(4)$$

Depending on the inclination of the foreground galaxy, different symmetries are
useful for scraping off the emission due to the foreground spiral in the
overlap region:
if the spiral is nearly face-on, rotation symmetry is used to
swing the unprojected portion around for subtraction; 
if the foreground spiral is instead more edge-on, its finite disk thickness 
may require reflection symmetry to be used to flip the 
unprojected portion of the spiral over for subtraction.
The opacities we determine directly are line-of-sight values, which
we correct to face-on values (assuming slab geometry). 
If the absorbing dust resides in spheres with only $\la1$ per line
of sight, then no correction is necessary.

                                \subsection{
Methodological Advantages and Caveats
                                }
In light of the ongoing controversy over selection effects in statistical
samples and the structural assumptions needed to interpret some of
the multiwavelength tests noted in the Introduction, 
there are several benefits to the direct, differential photometric approach 
we use to determine spiral opacities:
1) it is not subject to the selection effects which influence the statistical 
studies cited above;
2) there is no selection against high opacity regions, as there is in
some spectroscopic studies of small or point-like background objects 
shining through foreground disks ($e.g.$ quasars, or HII regions
in a background galaxy --- see James and Puxley 1993);
3) our imaging technique involves only differential photometry,
so calibration errors are not an issue;
4) large, contiguous areas can be analyzed, allowing average values of
the opacity to be estimated (whereas spectroscopic studies of background
HII regions or quasars probe relatively few points in a foreground disk
which induces a bias toward low absorption); 
5) there is no need to correct for the internal extinction of the 
background galaxy or the Milky Way (which is required in some spectroscopic
studies of background objects shining through foreground disks); 
6) our differential technique is not affected by color gradients due to
stellar population gradients (provided they are symmetric), which 
complicate the use of color maps ($V-K$, $B-K$, etc.) as dust detectors 
(Block et al. 1994);
7) we do not need to make assumptions about the vertical structure of the
disk or the relative $z$-distributions of stars and dust (these assumptions
are needed when the disk's own light is used to probe extinction  --- 
see Elmegreen 1980);
using the non-overlapping parts of the galaxies, we can test directly for
the requisite symmetry in azimuthal profiles; and
8) scattering corrections are also differential, which can keep them slight.

This technique also has some disadvantages relative to others:
1) there are rather few tractable objects nearby enough for spatially 
well-resolved analysis; and
2) the success of the technique hinges on the degree of symmetry in
both the foreground and background galaxies

The extinction values we derive differ in a significant way from 
those derived from internal galaxy properties.
Any technique relying on a galaxy's own radiation measures the dust content 
weighted by the distribution of starlight and dust in the galaxy itself. 
In our extinction measures, the light source (the background galaxy) 
is external to the opacity source (the foreground galaxy); thus, 
our extinction values are directly relevant, for example, to
calculations of the cumulative effect of spiral disks on optical quasar counts.
However, these measures may not be the most appropriate ones for calculating 
Tully-Fisher corrections, depending on the relative distributions 
of stars and dust.

It is useful to distinguish several regimes of galaxy backlighting,
depending on the apparent sizes of the galaxies and the impact
parameter of the background light. Quasars and distant galaxies
projected behind other galaxies represent a limiting case where the 
background sources are much smaller than foreground galaxies.  
These can provide reasonable extinction measures,
particularly since scattering into the beam is negligible,
but we may miss such small background objects if they are heavily absorbed
by the foreground galaxies --- that is, the use of such small
probes is biased toward the clearest lines of sight (e.g. Disney 1995).
Partially overlapping galaxies with similar angular size
are not vulnerable to this particular selection effect --- 
even extensive absorbing disks, almost 
opaque and larger than the luminous disk, would be detected in such pairs. 
Completely overlapping galaxies
(such as NGC 3314, Keel 1983) are good for probing the central
regions of disks, but with no $non$-overlapping parts to provide
estimates of the intrinsic disk brightness,
the results are necessarily limited in accuracy.
Finally, a spiral seen nearly edge-on may have its own disk backlit by its 
outer bulge (as analyzed by van Houten 1961 and more recently by Simien, 
Morenas, \& Valentijn 1993 and Knapen et al 1995). 
These cases have very well-understood geometry, but scattering effects can be 
much more important than in overlapping galaxy pairs (but were neglected in 
these studies). 
We will present our results on such ``peeking bulges" in a later paper.

				\subsection{
Corrections for Scattering Effects
				}

We estimate the possible role of scattered light in our measurements, with 
scattering acting to ``fill-in" extinction (Witt, Thronson \& Capuano 1992).
Scattering is potentially important because the bright central regions of 
a background galaxy may be close enough to the foreground galaxy for 
substantial amounts of light to be scattered into our line of sight 
by dust in the overlap regions.
Our technique automatically subtracts off internally scattered light in the 
foreground galaxy, to the extent that it is as symmetrically distributed
as the galaxy itself.
Our procedure is sensitive only to scattered light from the background galaxy. 
Furthermore, because we remove the symmetric counterpart of the foreground
galaxy from the overlap region, we are also removing some background
scattered light.
Thus, we are affected only by $differential$ scattering between 
the overlap region of interest and its symmetric foreground counterpart.
This differential scattered light drops very rapidly for increasing 
separation of the galaxies along the line of sight.  
Further details of our scattering corrections can be found in Appendix A.

                                \section{
Sample Selection                
                                }
Suitable partially overlapping galaxy pairs are rare; were it not for gravity
causing the galaxy covariance function to be much greater than unity 
at small separations, we would expect virtually no useful nearby candidates.
We attempted to find all overlapping galaxy pairs among galaxies bright 
enough and large enough (in angular size) for absorption measures. 
We examined candidate pairs on sky-survey images and obtained CCD images of 
the most promising.
 
Our observing sample is drawn from a variety of sources:
we performed numerical searches for overlapping neighbors in 
the {\it Revised Shapley-Ames} catalog (Sandage \& Tammann 1981), 
the {\it ESO-Uppsala} survey (Lauberts 1982), 
the {\it Uppsala General Catalog} (Nilson 1973), 
the {\it Revised New General Catalog} (Sulentic \& Tifft 1973),
the {\it NGC2000} catalog (Dreyer 1888, Sinnott 1988),
the {\it Morphological Catalog of Galaxies} (Vorontsov-Velyaminov \&
Krasnogorskaya 1962, Vorontsov-Velyaminov \& Archipova 1963, 1964, 1968, 1974),
the Karachentsev (1972) catalog of northern galaxy pairs,
the Chinese catalog of double galaxies (Zhenlong et al 1989) and the 
{\it Third Reference Catalogue of Bright Galaxies} 
(de Vaucouleurs et al. 1991).
We selected all pairs with center-to-center separations of less than 1.5 times 
the sum of their cataloged isophotal radii $R_{25}$, if such size information 
was present. 
We also selected individual catalog entries in the $UGC$, {\it ESO-Uppsala},
and $NGC$ listings which were typed as inherently multiple systems.
We also visually inspected all pairs in the Arp-Madore (1987) catalog 
(including all objects with notes mentioning dust or absorption),
the Arp (1966) {\it Atlas of Peculiar Galaxies}, 
and the Reduzzi \& Rampazzo (1995) catalog of southern pairs.
Further objects were drawn from visual inspection of the SRC J survey films in 
the Shapley Supercluster region.  Serendipity (while inspecting brighter 
candidates selected as above) and anecdotal lore provided a few more prospects, 
as well.
Throughout these searches, we were especially alert for any objects with clear 
signs of absorption, and we would certainly have selected any galaxies with 
extensive absorption appearing beyond the optical disk as ``bites" in 
background systems.

                                \section{
Observations                
                                }
Promising overlapping candidates were observed using CCD cameras at Kitt Peak, 
Cerro Tololo, and Lowell Observatory.
We have so far imaged 56 galaxy pairs, of which a dozen
are tractable enough for detailed analysis.  
An additional dozen may admit more limited analysis.
We have also observed several
``peeking bulge" galaxies: nearly edge-on spiral galaxies whose  
bulges can be seen on either side of their disks,
thus providing backlighting for the intervening disks; 
such individual systems can be analyzed in a similar,
but not identical, fashion as the overlapping pairs.
We concentrated on imaging in the $B$ and $I$ bands, to give the
quickest route to measures over a long color baseline without
risk of emission-line contamination.  We rejected the $U$ band for most 
objects, since the gain in wavelength baseline and the effective
extinction curve coverage is normally offset by losses due to fainter 
background light from early-type galaxies,
lower chip efficiency, and increased Poisson noise at a given flux level.

Most of our targets were observed with the 1.5m telescope at 
Cerro Tololo, using a Texas Instruments CCD binned during readout
to provide $400 \times 400$ pixels at 0.54 arcsecond/pixel
(in November 1992),
or a Tektronix $2048^2$ device giving 0.24 arcseconds/pixel
(July 1995).
Observations at the 1.1-m Hall telescope of Lowell Observatory 
(in March 1991) used
a Texas Instruments $800 \times 800$ CCD and 2:1 focal reducer,
covering a large field 9.4 arcminutes square at 0.708 arcsecond/pixel. 
This was especially important for pairs of large angular size such as 
NGC 4567/8 in Virgo.
At the 2.1m telescope of Kitt Peak National Observatory (June 1991), we 
used either the $1024^2$ Tektronix CCD at 0.19 arcseconds/pixel or
a STIS $1024^2$ chip at 0.27 arcseconds/pixel (with the detector switch
necessitated by a temporary detachment of the Tektronix chip from 
the cooling finger within the dewar).
 
Based on the CCD images, we
rejected many candidate pairs for being too asymmetric, for having 
the wrong galaxy in front (such as AM 0327-285; de Mello et al. 1995), for 
having a foreground star in the crucial region, or for morphological reasons 
(foreground E and S0 galaxies show no measurable absorption: $A_B\le0.1$
magnitude). 
The complete list of candidates imaged to date with the CCDs is given in 
Table 1, with reasons for the rejection of those not analyzed.

                                \section{
Analysis of Individual Objects
                                }
This paper reports the results for overlapping pairs which we have found to 
be most tractable.
Nonetheless, each system warrants 
individual discussion about the symmetry assumed or the particular
limitations suggested by its structure or geometry.
For the overlapping pairs analyzed in this work, their identifications, 
morphologies, isophotal radii and velocities are given in Table 2.  
In the following discussion of individual objects, we will tend to quote 
magnitudes of extinction $A$ rather than optical depths $\tau$, 
where $A=1.086\tau$.
Typical errors in individual measurements are $\approx0.15$ mag.
The objects are discussed in roughly descending order of quality,
but their results are tabulated in alphabetical order.

                                \subsection{
AM 1316-241
                                }
As reported in White \& Keel (1992), our best case thus far is AM1316-241, 
an Arp-Madore catalog object consisting of a foreground Sbc projected 
against a background elliptical.
Figure 2a shows the $B$-band image of this pair, which is also interesting
because the recession velocity of the foreground spiral (10365 km s$^{-1}$)
is 660 km s$^{-1}$ $larger$ than that of the background elliptical (the
single velocity listed in the ESO-LV catalog is attributed 
to the wrong pair member).
The axial ratio of the foreground spiral is 4.42, implying an inclination of
$77^\circ$.

Figure 2b shows a $B-I$ image, where the foreground overlapping spiral arm
very obviously reddens the light from the background elliptical.
The symmetry of each of the galaxies is good enough that we can employ
the image cut-and-paste technique described in \S2 to estimate the opacity over 
a relatively large fraction of the overlap region.  
Figures 2c-d show the resulting maps of $e^{-\tau^\prime}$ in the $B$ and $I$ 
bands, displayed with the same absolute intensity scale, the 
darker regions being more opaque.
The opacity is clearly concentrated in the 
spiral arm, while the interarm region is nearly transparent.
It is also obvious from Figures 2c-d that the arm is optically thicker in $B$ 
than in $I$. 
Table 3 lists the face-on-corrected extinction in the arm and 
interarm regions, as well as for an average over the disk area seen in the 
$e^{-\tau^\prime}$ maps of Figures 2c-d. 
In the ideal case of infinitely thin dust disks, the face-on-corrected 
extinctions are found by dividing the apparent 
extinction by the galaxy's axial ratio; this correction will be an overestimate
for the more realistic cases of finite thickness and clumped absorbers. 
The resulting extinctions are rather small: in the blue,
$A_B=0.38$ in the arm region and 0.08 in the interarm region, while in $I$,
$A_I=0.16$ and 0.05 in the arm and interarm regions, respectively. 
The arm is at 0.75 $R_{25}^B$ (where $R_{25}^B$ is the radius at which the
blue surface brightness $\mu_B=25$ mag arcsec$^{-2}$; see Table 2),
while the measurable disk region extends from $0.37-0.75$ $R_{25}^B$.
The radial extents of these various regions are also given in Table 3.

                                \subsection{
AM 0500-620
                                }
The E/Sbc pair AM 0500-620 shares some of the favorable characteristics of 
AM 1316-241 --- it is comprised of a relatively undisturbed foreground 
spiral and a symmetric background elliptical (see Fig. 3a for a $B$-band image).
In this pair as well, the elliptical can be accurately modeled from its 
unobscured half, and the spiral is symmetric enough for rotational symmetry to 
match its structure in some detail.  In practice, each galaxy was modeled
and subtracted from the data iteratively to converge on good models
for each component separately. The absorption follows the arm as traced
in $B-I$ quite closely (see Figure 3b). 
Along the arm ridge line, we find $A_B>3.0$ and
$A_I=2.1$, while the interarm extinction ranges over $A_B=0.1-0.6$ 
and $A_I=0-0.7$ at various points seen against the elliptical (see Table 3).
The symmetry of this system is good enough to allow the construction of
$e^{-\tau}$ maps, as for AM 1316-241 above.  Figures 3c and d show
maps of $e^{-\tau_B}$ and $e^{-\tau_I}$, respectively, with the same
absolute intensity scaling.

                                \subsection{
NGC 1738/9
                                }
Figures 4a and b are $B$ and $B-I$ images, respectively, of the Sbc pair 
NGC 1738/9.  
The symmetry of this system is not good enough to do an opacity analysis in 
the same detail as for AM 1316-241 and AM 0500-620.  
Instead, the two regions indicated in Figure 4a are investigated: a foreground 
arm region at 0.65 $R_{25}^B$ and an interarm region at 0.55 $R_{25}^B$.  
Symmetric regions in the foreground and background galaxies are used to infer 
the apparent extinction in $B$ and $I$ in these two regions.  
The apparent extinction values of the foreground galaxy (NGC 1739) are divided 
by its axial ratio of 1.95 (indicating an inclination of $59^\circ$) to give 
the face-on-corrected values listed in Table 3.
The face-on extinctions are again quite low: in the arm region, $A_B=0.3-0.37$ 
and $A_I=0.24-0.3$, while in the interarm region, $A_B=0.2-0.26$ and $A_I=0.16$.

                                \subsection{
NGC 4567/8
                                }
The Sbc pair NGC 4567/8 (UGC 7777/6) is another case where the analysis is 
limited by the general lack of symmetry (see Figures 5a and b for 
$B$ and $B-I$ images)
Here we concentrate on the dark lane in the upper left of Fig. 5a which
cuts across a brighter background galaxy arm.  
The comparison region for the foreground arm is taken from a region
along the arm but beyond the projected bulk of the background galaxy
(further to the upper left in Fig. 5a); the
comparison region for the background arm is along the background arm, just
away from where it is blocked by the foreground galaxy.
The foreground galaxy (NGC 4568/UGC 7776) has an axial ratio of 2.29, implying
an inclination of $64^\circ$.  
The assessed region in the foreground galaxy samples, in projection, a range 
of radii spanning $0.5-0.85$ $R_{25}^B$.
We calculate face-on extinctions of $A_B=1.1$ and $A_I=0.69$ for this 
region (see Table 3), which are substantially larger than in the previous
two systems.  

The interpretation of the light seen beyond 
the strong dust lane in NGC 4568 (to the lower left of the region 
analyzed above)
as coming from the background galaxy rather than foreground structure
hinges on whether any similarly bright areas are found at comparable 
projected radius in NGC 4568, and on the rather symmetric shape of
NGC 4567 as seen in the less-obscured $I$ band. Inspection of archival
HST ``snapshot" images obtained in the F606W filter (WFPC2 datasets
U29R4H01/2, PI G. Illingworth) shows that most of the excess light in this 
area comes from distinct bright clusters and associations, brighter than 
any others seen in the foreground object at comparable radius but quite 
comparable to the (systematically brighter) star-forming regions in the 
background system (see Fig. 5c). 
This somewhat strengthens our interpretation of the excess light
as indeed shining through a more transparent interarm medium.

The WFPC2 data also show that the darkest absorbing clouds in this pair, with 
a measured extinction of $A_{606}\approx 1.5$, are two irregular resolved
features about 7$^{\prime\prime}$ (0.5 kpc) in extent, but narrow enough 
($<1^{\prime\prime}$) that they
are not prominent in our ground-based images. Both are located well beyond the 
spiral arms (and other dust features) in
NGC 4568 (as marked on Fig. 5c). Their low residual intensity requires that 
they be in the foreground,
not part of NGC 4567 in the background. Either they are isolated in the outer 
disk, or are located several kpc from the disk plane (either of which might be 
attributed to the effects of interactions between these two galaxies). We cannot
immediately exclude the possibility that they are in the extreme foreground of 
the Milky Way itself, though the surface density of such clouds could not be 
very large without violating constraints from the number $not$ seen in HST 
imagery of elliptical galaxies, and the intensity of high-latitude IR cirrus 
emission.

                                \subsection{
UGC 2942/3
                                }
This is a pair of highly-inclined spirals, with the background galaxy
seen only a few degrees from edge-on (Figures 6a and b show $B$ 
and $B-I$ images).   The dust lane in the background galaxy
provides a recognizable target
to seek through the foreground disk. For cases like this, scattering
corrections become unnecessary since the edge of the background
dust lane is a sharp target; even small-angle scattering would 
contribute only over a larger angular scale.

To estimate the extinction in the foreground spiral UGC 2942, we
consider intensity slices perpendicular to the projected plane
of the background galaxy UGC 2943. These are taken in the
overlap region and at the corresponding locations on the opposite
side of UGC 2943. As shown in Fig. 6c, both $B$ and $I$ profiles show
a feature corresponding closely to the position and form of the
background galaxy, dimmed by factors of order 0.45 in B and 0.32 in $I$. 
The $B$ value is particularly uncertain due to foreground structure,
but even this detection is significant above the $3 \sigma$ level.
We can exclude the possibility that the ratio of $B$ and $I$ extinctions
follows a galactic reddening law, which most likely means that 
the extinction is dominated by a few regions of large optical
depth rather than widely-spread extinction. Some such structure is
visible in UGC 2942, especially in the I image (Fig. 6c). A foreground
dust lane crosses the northern part of the overlap region, and in
fact the signature of background light is detected only south of
this region. The implied optical depth across the spiral arm
(within the dust lane) is of order $\tau_B = 3$.

Both galaxies in this pair have reasonably widespread line emission, so that
measurement of a Balmer decrement from H II regions in the background
galaxy might give independent extinction measures for at least those
lines of sight where we detect background regions (as was done for
NGC 3314 by James \& Puxley 1993). We attempted such a
measurement for UGC 2942/3 using spectra obtained with the KPNO 2.1m 
telescope and GoldCam spectrometer, along the major axes of each
galaxy. Accurate emission-line rotation curves were measured (Fig. 6d), but 
the galaxies have orientations and rotation directions that defeat this 
technique; their rotation directions make the redshifts observed in the 
overlap region match to within a few tens of km s$^{-1}$.

                                \subsection{
AM 1311-455
                                }
AM1311-455 is comprised of a foreground ringed Sa projected against
a rather disturbed Sc/d galaxy.  
Dust in the resonance ring is clearly seen to attenuate light from the 
background galaxy in the $B$ image of Figure 7a.
Structure in the background galaxy is evident though the regions inside and 
outside the ring.
The resonance ring appears reddened in the $B-I$ color image of Figure 7b.  
In the ring itself (at $1.18R_{25}$) we find face-on corrected values of
$A_B\approx0.92$ and $A_I=0.31$.
We also analyzed regions to the interior of the ring, at $0.95R_{25}$, 
and find the disk to be nearly transparent: $A_B\approx0.17$ and  
$A_I\approx0.7$.

                                \subsection{
ESO 0320-51
                                }
ESO 0320-51 is a foreground, face-on ring galaxy projected against an 
edge-on S0 (see Figures 8a and b for $B$ and $B-I$ images).  
The ring galaxy is likely to have recently had a collision with the small 
galaxy seen projected just within the western edge of the ring.  
The $B$ image and the color ratio map show a slight discontinuity where the 
ring crosses the S0, which suggests the S0 is in the background.
Comparison of $B$ and $I$ images shows that the ring obscures the 
edge-on disk more in $B$ than in $I$, which more strongly indicated that
the S0 is in the background.
Detailed differential analysis confirms this, given that a small amount
of extinction is found in the ring (which lies at 0.65$R_{25}$: 
$A_B=0.3$ and $A_I=0-0.19$.
We also find a small amount of interarm extinction in $B$, $A_B\approx0.1$,
just within the ring at 0.5$R_{25}$; we find only an upper limit for the 
extinction in $I$: $A_I<0.1$.

                                \subsection{
NGC 3314
                                }
NGC 3314, a remarkable superposition of two spirals in the 
Hydra cluster (Abell 1060), was considered in the context of opacity 
measurements by Keel (1983). Our more recent imaging allows us to
greatly improve upon these measurements ($B$ and $B-I$ images are
shown in Figures 9a and b).
Following McMahon et al. (1992), we will call the foreground Sc galaxy 
NGC 3314a and the background Sab system NGC 3314b. Color-index maps,
the symmetry of rotation curves (Schweizer \& Thonnard 1985), and our $K$-band 
imagery show that the nuclei are separated by only 1.8$^{\prime\prime}$. 
We cannot do as complete an analysis here as for the best-case
partial overlaps with a background E/S0 galaxy, first because the
background object is a spiral (albeit of early type) and second because
the overlap is so nearly central that there is no empirical check on
the brightness profile of the background galaxy. However, this system is 
uniquely valuable because we can estimate extinctions in the foreground 
galaxy closer to its center than in any other of our sample.

The best places for reliable extinction measurements are the points where
the arms of NGC 3314a cross the disk edges of NGC 3314b, going from
projection against the bright disk to projection against almost
blank space at essentially the same radial distance for the arm.
We measured the arm intensities at adjacent points on and off the 
background disk, after subtracted a minimal exponential-disk
model to flatten most of the background gradient (so that interpolation
to get the relevant unobscured background intensity is better constrained).
For two locations where the arms cross the disk at about 0.5$R_{25}$, 
both $A_B$ and $A_I$ are comparable at 1.8, while the interarm
regions average $A_B=0.60$ and $A_I=0.34$.

The H I maps presented by McMahon et al. (1992) afford a chance for a crude
measurement of the dust-to-gas ratio, as represented by $N(H I)/E_{(B-V)}$, 
limited by the resolution of their VLA H I synthesis (FWHM about 
14$^{\prime\prime}$, as compared to the arm width (traced by optical 
extinction) of about 5$^{\prime\prime}$
in the regions we have analyzed. Their Figs. 3 and 4 suggest a column
density on H I of about $10^{21}$ cm$^{-2}$, and application of the
usual Galactic extinction law to our values of $A_B$ implies a ratio
$N(H I)/E_{(B-V)}> 2 \times 10^{21}$ cm$^{-2}$; the upper limits is due
to the likelihood that the H I is clumped into arms not well resolved in
the H I map. Thus we find a ratio of the same order as in our local 
neighborhood, and it is not clear how close a correspondence we should
expect even for identical grain populations due to the effects of unresolved
clumping on the spatially-averaged extinction values we measure.

                                \subsection{
NGC 450/UGC 807
                                }
This galaxy pair is comprised of NGC 450 (UGC 806), an Sc/Sd system at 
$cz=1863$ km s$^{-1}$, and UGC 807, a spiral of earlier type at $cz=11587$ 
km s$^{-1}$ (Figures 10a and b show $B$ and $B-I$ images).  
Rubin \& Ford (1983) sought luminosity and distance indicators from 
rotation curves of this pair (with conclusions disputed by Moles et al. 1994).
The large redshift difference effectively rules out the possibility
of interaction, so that the line-of-sight distance is large
and scattering effects can be ignored. We used two approaches
to remove the foreground light from NGC 450. One parallels that
used by Andredakis \& van der Kruit (1992) for this pair ---
modeling the whole foreground galaxy with the IRAF {\it ellipse} task,
letting it average over small-scale structure, and subtracting the
resulting smooth model. Since UGC 807 has a substantially
smaller angular diameter than NGC 450, we could also make radial cuts
adjacent to it and interpolate between them as a more local measure
of foreground light. In neither case do we detect any extinction upon
comparison of the ``inner" and ``outer" halves of UGC 807
in surface brightness or color, to limits of $\Delta (B-I) < 0.05$ and 
$A_B < 0.1$ across the outer disk edge. The measured area lies at about
0.95---1.0$R_{25}$.

                                \subsection{
NGC 4647/9
                                }
NGC 4647/9 is a bright, well-known pair in the Virgo cluster
(see the $B$ band image in Figure 11). NGC 4647 (UGC 7896) is a spiral
with flocculent structure and heavy dust lanes, especially prominent
at the edge of the optical disk (see, for example, the photograph in
Arp 1966, where this pair is number 116). They are projected at the large
center-to-center separation of 11.9 kpc 
(for a distance of 16 Mpc) even compared to the
large scale of the elliptical NGC 4649 (= M60 = UGC 7898), so this pair offers 
a chance to examine primarily any dust structure which might lie
beyond the bright optical
disk (since the entire spiral is projected against detectable light
from NGC 4649). The elliptical was modeled in two stages, using the
$STSDAS$ ``ellipse" task for the inner parts of the galaxy and
the best-fit global $r^{1/4}$ model beyond $r=106$ arcseconds,
to avoid the spiral's influence on fitted isophotes. The outer
regions are fitted by a somewhat shallower profile ($R_e=82^{\prime\prime}$)
than the global value of 68$^{\prime\prime}$ listed in the RC3
(de Vaucouleurs et al 1992). After subtraction of this model for
the elliptical component NGC 4649, no absorption structure is
found beyond the edges of the disk of NGC 4647, with the outermost
detected absorption associated with the prominent dust lane on the
northern side of the disk. {\it If} the spiral
is in fact in front, no dust features large enough to resolve have 
$A_B$ or $A_B>0.15$. The range sampled here is at and outside $R_{25}$.

                                \section{
Summary and Discussion
                                }

We have presented absolute extinction measures for ten spiral galaxies
in overlapping pairs. 
For each pair, there is some range of radii for which we can measure the 
residual intensity of background light transmitted through the 
foreground disk. 
We translate these measures into arm and interarm extinctions
(where such a distinction is possible) in both $B$ and $I$ bands.
In almost all cases, there is a large difference between arm and interarm 
values. 
In arm regions, $A_B\approx0.3-2$ and $A_I\approx0.15-1.6$, while 
in interarm regions, $A_B\approx0.08-1.1$ and $A_I\approx0.05-1.6$.  
Table 3 summarizes the pairs and regions for which extinction 
measurements have been made. 
Figure 12a graphically summarizes these results for arm regions,
while Figure 12b does the same for interarm regions
(solid and dotted diamonds represent values of $A_B$ and $A_I$, respectively).
The arm and interarm plots are drawn to the same scale to emphasize
that arm regions tend to be much more opaque than interarm regions.
Within each plot it is also clear that there is more extinction in $B$ than
in $I$, as expected.

The interarm (``disk") extinction tends to decline with radius (Fig. 12b) 
from $A_B$ values of only $\sim1$ magnitude within $\sim0.3R_{25}$. 
In contrast, spiral arms and resonance rings can be optically thick at
almost any galactocentric radius. 
We do not see evidence for substantial extinction in the 
outer parts of disks (and such extinction would have been obvious even 
in our initial screening as ``bites" taken out of background galaxies).
If we fit a single exponential to the distribution of extinction
in our ensemble, we get a scale length
$h_d\approx 0.3 R_{25}$ in both bands $B$ and $I$. Using the data from
Simien \& de Vaucouleurs (1986), typical spiral disks have a stellar 
exponential disk scale length of $h_s\approx 0.28 R_{25}$. We thus find 
that the interarm dust has the same scale length as the disk starlight,  
in agreement with the Kylafis \& Bahcall (1987) result of near--equality 
found from photometric decomposition of the edge-on spiral NGC 891.  
Presently known sites of grain formation --- in particular kinds of stellar 
atmospheres and expanding envelopes --- naturally give rise to
dust distributions which are tied to those of stars.

Our initial results on AM1316-241 (White \& Keel 1992)
led us to conclude that disk opacity is 
concentrated in spiral arms and that interarm regions are fairly transparent.
Our newer work is generally consistent with this picture, with resonance
rings found to be as optically thick as spiral arms.
Therefore, the distribution of absorption tends to be spatially
correlated with particularly bright regions, 
since spiral arms are brighter than interarm regions.
We suggested (White \& Keel 1992) that this spatial correlation between 
internal extinction and emission may account for the statistical results 
reported in earlier studies --- that surface brightness is roughly 
independent of inclination. The dust is optimally placed to affect 
global blue photometric properties, since typically half the disk light
comes from only about 20\% of its area, accounting for the rather
flat inclination-surface brightness relation, $without$ requiring 
galaxies to be optically thick in interarm regions.
These remarks are directed more to grand-design spirals, since in flocculent 
systems we cannot make a clear distinction between arm and interarm regions.

Closer examination of Table 3 shows that for some of the galaxies, the 
disparity between $A_B$ and $A_I$ is not as great as that in our own galaxy, 
which has $A_B/A_I\approx2$.
Several of these galaxies have $A_B/A_I \approx 1.5$, so their
extinction curves are flatter (``greyer") than the Galactic curve.
Since our measurements are based on spatially averaged transmission
values, the ``effective" extinction may not be fully comparable to
the extinction curves derived from what are essentially point sources
in our own and nearby galaxies. 
In particular, since the dust distribution is directly observed 
to be clumpy on a wide range of scales, we may expect to see such a 
flattening of the observed extinction compared to that which would
come from the intrinsic grain properties. 
Clumped extinction will saturate in $B$ before $I$, which diminishes $A_B/A_I$.
The more strongly clumped the dust, the flatter the extinction curve will be.
In viewing a spatially extended region, the light at each 
wavelength comes preferentially from the areas with smallest extinction. 
As a simple example, if we
consider a uniform dust screen with transparent holes which occupy some
covering fraction $\eta$, the measured extinction curve from an extended
background source will never give an effective extinction greater than 
$A=-2.5 \log \eta$ regardless of the optical depth in the screen;
that is, if 10\% of the area has no extinction, at no
wavelength would we measure an extinction greater than 2.5 magnitudes.
We expect some conceptually similar (but naturally much more
complex) situation to obtain in the disks of spirals.
Our limited sample does not show any systematic difference in the
slope of the effective extinction between arm and interarm
regions, or any overall trend with radius within the galaxies.
Further observations, particularly $HST$ imaging to trace the
extinction structure to scales of order 10 pc and $ISO$ measurements 
to measure the overall dust masses, are scheduled to examine the role
of clumpiness in more detail.

Our results bear on the question of whether the high-redshift ``QSO cutoff"
can be produced by absorption in spirals along the line of sight.
The high redshift of the cutoff offers plenty of room for even modest 
individual optical extinctions to have an impact, particularly if 
the effective extinction curve rises as steeply in the
UV as the Galactic extinction curve does. For a fiducial set of
spiral galaxy parameters,
Ostriker \& Heisler (1984) estimate that 50\% of QSOs at $z=4.5$ will suffer 
such obscuration by foreground galaxies; this is close 
enough to the characteristic peak redshift in the QSO distribution at 
$z \approx 2.2$ to make obscuration effects worth investigating.
We find that disks are optically thin in spiral types Sb and later, 
which have $A_B < 1$ from 0.5--0.9 $R_{25}$; extinctions are below our 
measurement errors for $R>R_{25}$. 
The typical $inter$arm behavior of our sample is very
close to the model adopted by Ostriker \& Heisler. 
Their fiducial model is based on the radial structure of the
Milky Way and the integrated extinction perpendicular to its disk at the
solar location $R_\odot$.  Since we give our results in terms
of $R/R_{25}$, we make contact with their results 
by noting that $R_\odot/R_{25}=$ 8 kpc/11.5 kpc = 0.7 (following 
de Vaucouleurs \& Pence 1978). This implies that the Ostriker \& Heisler 
model has $A_B=0.9$ at 0.5$R_{25}$, which is quite consistent with the 
interarm $A_B$ values (typically near unity) we find at this radius. 
Spiral arms will provide additional absorption, but they cover much less 
than half of the surface area in grand-design spiral disks.  The covering 
fraction of spiral arms tends to be larger in flocculent spiral galaxies, 
however.  Even given the uncertainties in the relative demographics of 
grand-design and flocculent spirals, the cumulative opacity from
spiral galaxies is unlikely to be more than a few times larger than the 
fiducial model adopted by Ostriker \& Heisler,
especially if the dust content declines with increasing redshift
(the dust content probably grows with cosmic time due 
to continuing production in stars). Thus, the accumulation of spiral 
disks in the line of sight can reduce QSO counts by $>50\%$, but not 
enough to induce the QSO ``cutoff."

We will report elsewhere on our studies of ``peeking bulge" systems, 
in which the bulge of a nearly edge-on spiral backlights part
of its own disk, since 
their analysis is more subtle.  To avoid underestimating the optical depths
in the intervening disks, one must be sure to scrape off the emission
due to the intervening disk, which is difficult to estimate from the 
symmetric regions on the far side;
furthermore, such systems are likely to have forward-scattered bulge light 
``fill in" much of the true absorption. 
Future papers in this series also include extension of the extinction curve
coverage for some of these overlapping galaxies to $U$ and $K$ bands, and 
the use of slit spectroscopy in overlapping regions to determine photon 
ownership by exploiting Doppler shift differences between overlapping galaxies. 

				\acknowledgments

We thank Barry Madore for providing an electronic version
of the AM catalog, greatly simplifying part of our search task. 
Duilia de Mello kindly obtained ESO redshifts of AM1316-241 on our behalf.
Mayo Greenberg and Adolf Witt updated our knowledge of the 
scattering parameters for grains.
We acknowledge support from EPSCoR grant EHR-9108761 and NSF
REU grant AST-9424226. We thank the director of Lowell Observatory
for time to observe some of the pairs of especially large angular
size, which formed an important complement to our KPNO and CTIO imaging.
The Digitized Sky Survey was produced at the Space Telescope Science Institute
under U.S. Government grant NAG W-2166. The images of these surveys are based
on photographic data obtained using the Oschin Schmidt Telescope on Palomar
Mountain and the UK Schmidt Telescope. The plates were processed into the 
present compressed digital form with the permission of these institutions.
The HST archival images were retrieved from the Space Telescope Science 
Institute, which is operated by AURA, Inc., under NASA contract No. NAS5-26555. 

                                \appendix
				\section{
Scattering Corrections}

We attempt to calculate a maximum role for scattering as follows.
We take the major-axis profile of the background galaxy, and
assume the galaxy to be circular with this profile as seen from
each point in the foreground system. We further assume the dust
to be uniformly distributed, as this is the most effective way
to scatter light from a fixed amount of dust. Taking the geometry
shown in Fig. 13, we numerically integrate the intensity of
scattered light as a fraction of the transmitted light, both
normalized to the background intensity at the overlap position.
We use the Henyey-Greenstein (1941) expression for the phase function
during scattering, which becomes 
$$ {{dI(r,\theta)} \over {I}} = {{ \tau a (1-g^2)} \over {(1+g^2-2g \cos 
\theta)^{1.5}}}$$
in current notation; here $I$ is the intensity of the background 
galaxy in the direction specified 
by $r$ and $\theta$, $\tau$ is the scaling by optical depth, $a$ is the 
albedo at the relevant wavelength, and $g$ is the asymmetry parameter.
Based on the work of Witt et al. (1990, 1992) and Calzetti et al. (1995),
we use $a=0.6$ at both $B$ and $I$, and $g=0.8$.
Since the line-of-sight separation is not directly known, we allow this
to vary over the entire plausible range. For example, if neither pair
member is morphologically distorted, the two galaxies probably do not
physically overlap at the relevant radius. Fig. 14 shows a sample
calculation of scattering intensity for AM 0500-620. Both the total
and differential scattering are shown on a logarithmic intensity scale,
dropping rapidly with assumed separation as the background galaxy
occupies a decreasing solid angle as seen from the scatterers.
For galaxies more than a few radii apart, the effect becomes negligible
(so we can ignore it for galaxies with very different redshifts).
Scattering redistributes radiation in angle over the
characteristic scale of the phase function, so that if one traces
sharp features such as dust lanes, scattering will not affect the
small-scale structure. This means that scattering is not important in those 
pairs where we use dust-lane or arm edges as the background tracers, as 
in UGC 2942/3 and NGC 3314.

We tabulate in Table 4 the adopted minimum plausible
line-of-sight separation between galaxies for pairs in which scattering might
be an issue, based on the outermost 
symmetric isophotes, and the maximum corrections for differentially
scattered light at this separation. The table lists the projected
distance between the innermost overlap region and the background nucleus,
the minimum plausible line-of-sight separation between galaxies in units of 
this projected separation, and the calculated maximum differential
scattered intensity as a fraction of the unabsorbed background
light at the overlap location.
The relative corrections (scaled to unit optical depth $\tau$) are the
same at $B$ and $I$, since we adopt a constant albedo across this
wavelength range. As is apparent from the values in the table,
the maximum corrections due to scattering are always less than a few
percent in residual intensity, so that this is not a major
uncertainty in our results.

                                \clearpage
                                \title{
Figure Captions			}

                                \figcaption{
A cartoon of the ideal galaxy pair for our analysis. The light from stars in the
foreground and background systems is denoted by $F$ and $B$; their values in 
the overlap regions are estimated from the values $F^\prime, B^\prime$
in symmetrically located regions on the non-overlapping sides of the
galaxies.
				}
                                \figcaption{
AM 1316-241: a) $B$-band image; b) $B-I$ color image; 
c) $e^{-\tau_B}$; d) $e^{-\tau_I}$, with the
$e^{- \tau}$ images rotated to align with the spiral's major axis. Both galaxies
in this pair are symmetric enough to allow the detailed decomposition illustrated
in Fig. 1.
The strong absorption is concentrated into the projected spiral arm, with
much less in the interarm region just inside it. The opacity maps are
displayed at the same brightness scale, showing how much smaller the
extinction is at $I$ compared to $B$.
				}
                                \figcaption{
AM 0500-620: a) $B$-band image; b) $B-I$ color image; 
c) $e^{-\tau_B}$; d) $e^{-\tau_I}$. The box in the $B$ image 
shows the area enlarged in the residual intensity $e^{-\tau}$ maps. North is at 
the top and east to the left. The dust arm crosses from the lower left
corner to the right center edge. What appears to be a very red foreground star
appears just to the south of this arm, most apparent in the $B-I$ image.
Again, the scaling for the opacity images is identical for $B$ and $I$.
				}
                                \figcaption{
NGC 1738/9: a) $B$-band image; b) $B-I$ color image. In this pair of spirals, 
averages were
taken over the marked regions, and symmetric areas (with regard for the spiral
pattern) were used to estimate both foreground and background contributions.
				}
                                \figcaption{
NGC 4567/8: a) $B$-band image; b) $B-I$ color image; c) mosaic of WFPC2 F606W 
images, rotated to
the nearest quadrant from cardinal orientation. The HST imagery shows the
brightest associations used to attribute light past the dusty arm of
NGC 4568 to the background arm of NGC 4567. This mosaic also shows
several narrow dust clouds of high optical depth beyond the main disk
of NGC 4568 (within the dashed circles). They are sufficiently smeared
by seeing to be inconspicuous in the Lowell image above.
				}
                                \figcaption{
UGC 2942/3: a) $B$-band image, logarithmically scaled; the white rectangles 
show the areas averaged for the intensity strips compared in part 6c.
b) $B-I$ color-ratio image; 
c) (left two panels) Intensity slices parallel to the minor axis of UGC 2943, 
crossing the overlap (solid) and symmetric (mirrored, dashed) locations.
The vertical bar indicates the deepest part ofthe dust lane in
the background system UGC 2943, and the amount of extinction is
measured from the relative intensity depth of this dip in the
two slices at each passband. A region 10$^{\prime\prime}$ wide was averaged
for each intensity trace.
d) (right two panels) H$\alpha$-[N II] emission-line rotation curves for 
UGC 2942/3. Error bars are $\pm 2 \sigma$ from photon statistics, and the 
lower curves trace the
red-light intensity along the slit. The near coincidence of radial
velocity in the overlap regions defeated our attempt to use redshift
separation to distinguish emission from the individual galaxies. Radial
velocities are shown in the heliocentric frame; we derive nuclear
redshifts of $6261 \pm 5$ km s$^{-1}$ for UGC 2942 and $6269 \pm 20$
for UGC 2943.
                                }
                                \figcaption{
AM 1311-455: a) $B$-band image; b) $B-I$ color image. In the color-ratio image, 
differences in seeing 
between the $B$ and $I$ data have been largely compensated by smoothing the
$I$ image with the best-matching Gaussian. Note that features in the arms
of the background Sc system can be traced across the resonance ring
in the SBa foreground galaxy, confirming low extinction immediately
within the resonance ring.
				}
                                \figcaption{
ESO 0320-51: a) $B$-band image; b) $B-I$ color image. The combination of 
color and intensity data
indicate that the edge-on galaxy is in the background, thus probing
the ring and disk of the foreground, face-on system.
				}

                                \figcaption{
NGC 3314: a) $B$-band image; b) $B-I$ color image. Extinction measurements 
in this pair used slices
along the foreground arms on and off the dust lanes, and the amplitude of
the disk edge from the background galaxy as transmitted, to yield opacity
estimates. The superposition is almost perfect in this instance, with the
nuclei separated by only 1.8".
				}

                                \figcaption{
NGC 450/UGC 807: a) $B$-band image; b) $B-I$ color image. No reddening or 
extinction was detected in 
this pair. The smaller galaxy UGC 807 has a redshift six times as great as
NGC 450 and is thus clearly in the background.
				}
                                \figcaption{
NGC 4647/9: a) $B$-band image; b) $B-I$ color image. 
There remains ambiguity in this pair as to whether
the spiral is in front or behind, since both are Virgo cluster members and
we detect no absorption outside the spiral disk against the extensive 
envelope of the elliptical NGC 4649.
				}
                                \figcaption{
Summary of all face-on-corrected extinction measurements. a) (left)
extinction magnitudes in arm regions as function of $R_{25}^B$; 
b) (right) extinction magnitudes in inter-arm
regions as function of $R_{25}^B$. The arm measurements show no obvious
trend with galactocentric radius, but the interarm extinction drops
with distance from the nucleus in a way that can be well described by 
an exponential in extinction (and thus in column density). The scale length
of this form is close to that for the disk starlight in a typical spiral.
				}
                                \figcaption{
Schematic diagram showing the geometry and coordinate system used for
calculating scattering correction. The relevant angle $\theta$ in the 
scattering phase function is evaluated between the projected line of sight 
into the background system and each point in the background galaxies,
centered (as shown) on the point in the foreground system at which
the line of sight passes through its disk -- the location at which
we are measuring the extinction. The center of the background galaxy
is in the plane of the coordinate grid. Our numerical estimates
assume a uniform dust screen in the foreground galaxy, which is the
most effective configuration for scattering into the line of sight and
thus furnishes an upper limit to the possible correction for scattered
light.
				}
                                \figcaption{
Sample plot of relative scattering intensity versus assumed line-of-sight
separation between the galaxies for AM 0500-620, calculated using
the $B$ profile of the background elliptical. Both total and differential
scattering contributions are shown, to indicate how rapidly the
differential correction (to which our technique is sensitive) drops with
distance between the galaxies. The structure in the differential-scattering 
curve reflects purely
numerical fluctuations associated with the spacing of grid points
with respect to radii at which the surface-brightness profile is
tabulated. The outermost symmetric isophotes
suggest that the minimum allowed separation in the line of sight is
8 times the radius of the overlap region from the foreground nucleus.
Even for the minimum plausible separation, the maximum
role for scattered light is well within the errors of our extinction
measurements.}

\newpage

\begin{deluxetable}{lcl|lcl}
\tablecaption{Candidate Overlapping Galaxy Pairs}
\tablewidth{0pt}
\tablecolumns{6}
\tablehead{
\colhead{Pair} & 
\colhead{Observatory} & 
\colhead{Notes} & 
\colhead{Pair} & 
\colhead{Observatory} & 
\colhead{Notes}
}
\startdata
AM 0247-312  & CTIO & S0+E	     & NGC 450      & CTIO  & this paper \nl
AM 0313-545  & CTIO & interfering star & NGC 1531   & CTIO  &tidal arm overlap\nl
AM 0327-285  & CTIO & S behind       & NGC 1738/9   & CTIO  & this paper \nl
AM 0500-620  & CTIO & this paper     & NGC 2207     & KPNO  & possible \nl
AM 0546-253  & CTIO & two SBs	     & NGC 3314     & CTIO  & this paper \nl
AM 0645-264  & CTIO & tidal dist.    & NGC 4567/8   & Lowell & this paper \nl
AM 1311-455  & CTIO & this paper     & NGC 4647/9   & KPNO  & this paper \nl
AM 1316-241  & KPNO & this paper     & NGC 5090/1   & CTIO  & asymmetric S \nl
AM 2030-303  & CTIO & irr. structure & NGC 5544/5   & Lowell & possible \nl
AM 2131-572  & CTIO & interfering star  & NGC 6050     & KPNO  & possible \nl
AM 2344-282  & CTIO & pair 1 - too small? & NGC 7016& CTIO  & possible \nl
AM 2344-282  & CTIO & pair 2 - too small? & NGC 7119& CTIO  & interfering star\nl
AM 2347-292  & CTIO & too small?     & NGC 7174     & CTIO  & tidal dist. \nl
AM 2354-304  & CTIO & SBb+Sb   	     & NGC 7284/5   & CTIO  & E in bkgnd \nl
Anon 2345-29 & CTIO & S+S, faint     & NGC 7433     & CTIO & possible \nl
Arp 40       & KPNO & possible       & UGC 2942/3   & CTIO  & this paper\nl
ESO 0245-53  & CTIO & possible       & UGC 3445     & Lowell & too distorted \nl
ESO 0320-51  & CTIO & this paper     & UGC 3995     & Lowell & possible \nl
ESO 0416-50  & CTIO & possible       & UGC 4619     & Lowell & possible \nl
ESO 0433-41  & CTIO & inclined S+S   & UGC 7535     & CTIO & possible \nl
HCG 5        & CFHT & from P.Hickson & UGC 8813     & Lowell & S0+S0 \nl
IC 4378      & CTIO & possible       & UGC 8972     & KPNO  & possible \nl
IC 4721      & CTIO & possible       & UGC 9554     & CTIO & possible \nl
IC 5328      & CTIO & E+S0	     & UGC 10049    & KPNO & possible \nl
IC 5349      & CTIO & S0+compact     & UGC 10422    & Lowell & possible \nl
IC 5364      & CTIO & possible       & UGC 11168    & KPNO & too detached \nl
MCG 2-58-11  & CTIO & late-type S+S  & Zh0016-61    & CTIO & possible \nl
NGC 45       & CTIO & small bkgnd group & Zh2222-31 & CTIO & possible \nl
\enddata
\end{deluxetable}

\newpage

\begin{deluxetable}{lccc|lcc}
\tablecaption{Overlapping Galaxy Pair Properties}
\tablewidth{0pt}
\tablecolumns{7}
\tablehead{
\colhead{Foreground} & 
\colhead{type}	& 
\colhead{$cz$}	& 
\colhead{$R_{25}^B$} & 
\colhead{Background}	& 
\colhead{type} 	&
\colhead{$cz$} \nl
\colhead{ {\it alternate name} }&	
\colhead{} & 
\colhead{(km/s)} & 
\colhead{(arcsec)} & 
\colhead{ {\it alternate name} } & 
\colhead{}&
\colhead{(km/s)}
}
\startdata
AM 0500-620	& Sbc	& 8420	& 28 & ESO-LV 1190271& E	& 9200  \nl
ESO-LV 1190270	&	&	& &  		  &     &   \nl
 & & & & & \nl
AM 1311-455	& Sa	& 3091	& 60.0 &		& Sc	& 3110 \nl
 & & & & & & \nl
AM 1316-241	& Sbc	& 9554	& 37.5 & ESO-LV 5080450 & E 	& 4317  \nl
ESO-LV 5080451 &	&	& 	&	 &	& \nl
 & & & & & \nl
ESO 032012-5150.1 & Sa	& 17328	& 22.5 &		& S0	& \nl
FAIRALL 299	&	&	&	&	&	& \nl
 & & & & & & \nl
NGC 450		& S	& 2118	& 92.7 & UGC 807	& S	& 11431 \nl
UGC 806		&	&	& 	&	&	& \nl
 & & & & & \nl
NGC 1739	& Sbc	& 3892	& 42.0 & NGC 1738	& Sbc	& 3978  \nl
 & & & & & & \nl
NGC 3314a	& S	& 2872	& 46.5 & NGC 3314b	& S	& 4426 \nl
 & & & & & & \nl
NGC 4568	& Sbc	& 2255	& 137.0 & NGC 4567	& Sbc	& 2274 \nl
UGC 7776	& 	& 	& 	& UGC 7777 	& 	& \nl
 & & & & & \nl
NGC 4647	& S	& 1422	& 86.5 & NGC 4649	& E	& 1413 \nl
UGC 7896	& 	&	& 	& UGC 7898 / M60 & 	& \nl
 & & & & & \nl
UGC 2942	& S	& 6361	& 39.5 & UGC 2943	& S	& 6434 \nl
\enddata
\end{deluxetable}

\newpage

\begin{deluxetable}{c|ccc|ccc|ccc}
\tablecolumns{10}
\tablewidth{0pt}
\tablecaption{Face-on Extinctions}
\tablehead{
\colhead{} & 
\multicolumn{3}{c} {--------------- arm ---------------} & 
\multicolumn{3}{c} {------------ interarm ------------} & 
\multicolumn{3}{c} {--------- average ---------} \nl
\colhead{Galaxy} 
& \colhead{$R/R_{25}^B$} & \colhead{$A_B$} & \colhead{$A_I$} 
& \colhead{$R/R_{25}^B$} & \colhead{$A_B$} & \colhead{$A_I$}
& \colhead{$R/R_{25}^B$} & \colhead{$A_B$} & \colhead{$A_I$} 
}
\startdata
AM 0500-620	& 0.6		& $>$2.3	&  1.64    
		& 0.5  		&  0.1-0.47 	& 0.0-0.55  
		&         	&      		& 		\nl
AM 1311-455 	& 1.18     	&  0.73     	&  0.24    
		& 0.95 		&  0.17 	&  0.07 
		&          	&      		& 		\nl
AM 1316-241 	& 0.75     	&  0.38     	&  0.16    
		& 0.4-0.75	&   0.08    	&   0.05    
		& 0.4-0.85 	& 0.19 		& 0.15 		\nl
ESO 0320-51 	& 0.65     	&  0.27     	&  0.17    
		& 0.50 		&   0.1 	&   $<$0.1 
		&          	&      		&  		\nl
NGC 450     	&          	&          	&          
		&        	&           	&           
		& 0.95-1.0 	& $<$0.1 	& $<$0.1 	\nl
NGC 1739    	& 0.65     	& 0.30-0.37 	& 0.24-0.3 
		&    0.55 	& 0.20-0.26 	&    0.16  	\nl
NGC 3314    	&  0.16    	&  1.60     	&  1.24    
		&    0.19 	&   1.11    	&   1.60    
		&          	&      		& 		\nl
            	&  0.34    	&  1.64     	&  0.82    
		&    0.28 	&   0.77    	&   0.59    
		&          	&      		& 		\nl
            	&  0.42    	&  1.11     	&  0.82    
		&    0.39 	&   1.75    	&   0.63    
		&          	&      		& 		 \nl
NGC 4568    	& 0.5-0.85 	&  1.1      	&  0.69    
		&         	&           	&           
		&          	&      		& 		\nl
NGC 4647	& $\ge1$	& $<$0.15 ?	&   $<$0.15 ?
		&		&		&
		&		&		&		\nl
UGC 2942    	&          	&           	&          
		&         	&           	&           
		&     0.56 	& 0.35 		& 0.32 		\nl
\enddata
\end{deluxetable}

\clearpage

\begin{deluxetable}{lccc}
\tablecolumns{4}
\tablecaption{Scattering corrections for overlapping pairs}
\tablewidth{0pt}
\tablehead{
\colhead{} & 
\colhead{Projection}	& 
\colhead{Minimum separation/}	& 
\colhead{Scattering} \nl
\colhead{Pair}	& 
\colhead{radius}	& 
\colhead{projection radius}	& 
\colhead{fraction}
}
\startdata
AM 0500-620 &  \phn 6$^{\prime\prime}$ & 9.0  & 0.03\phn  \nl
AM 1316-241 &  \phn 6$^{\prime\prime}$ & 8.5  & 0.01\phn  \nl
AM 1311-455 & 52$^{\prime\prime}$ & 3.6  & 0.05\phn  \nl
NGC 1738/9  & 18$^{\prime\prime}$ & 3.7  & 0.027 \nl
NGC 4567/8  & 39$^{\prime\prime}$ & 5.5  & 0.04\phn  \nl
\enddata
\end{deluxetable}

\end{document}